\documentclass[12pt]{iopart}
\expandafter\let\csname equation*\endcsname\relax
\expandafter\let\csname endequation*\endcsname\relax
\usepackage{amsmath}
\usepackage{amssymb}
\usepackage{graphicx}
\usepackage{lineno}
\usepackage{color}
\usepackage{caption}
\usepackage{subcaption}
\usepackage{bm}
\usepackage{setspace}
\usepackage{hyperref}

\begin{document}
\title[On the electromagnetic effects of collisionless trapped-electron modes]{On the electromagnetic effects of collisionless trapped-electron modes}

\author{Yao Yao$^{1}$, Haotian Chen$^{1*}$, Yang Chen$^{2}$, Jiquan Li$^{1}$ and Xuru Duan$^{1}$}

\address{$^1$ Southwestern Institute of Physics, Chengdu 610041, China}
\address{$^2$ Department of Physics, University of Colorado at Boulder, Boulder 80309, USA}

\ead{chenhaotian@swip.ac.cn}

\begin{abstract}
We present a linear gyrokinetic theory for the electromagnetic collisionless trapped-electron mode (CTEM).
It is found that the weak electromagnetic effects of CTEMs originate from the particle dynamics.
Theoretical analysis reveals that the kinetic and fluid-like components of the trapped-electron parallel current cancel at leading order.
The ion parallel current is also negligible due to the weak ion transit resonance.
Consequently, the perturbed parallel current in the electromagnetic CTEM is dominated by passing electrons.
We demonstrate that these characteristics of particle dynamics decouple the CTEM from the shear Alfv\'en wave branch, rendering the electromagnetic effects subdominant.
Both eigenmode analyses and gyrokinetic simulations validate these findings.
\end{abstract}

\vspace{2pc}

\section{Introduction}
Anomalous transport driven by microturbulence is a crucial obstacle to achieving high energy confinement in tokamaks~\cite{Horton1999}.
At the ion gyroradius scale, the ion temperature gradient (ITG) mode and the collisionless trapped-electron mode (CTEM) play a dominant role in driving anomalous heat and particle transport.
Since future fusion reactors are expected to operate in high-$\beta$ regimes to increase fusion power density and economic efficiency, understanding the electromagnetic effects of these microinstabilities is essential.
The ITG mode is known to be stabilized by electromagnetic effects through the coupling between the drift wave and the shear Alfv\'en wave (SAW) branch~\cite{Kim1993}.
However, gyrokinetic simulations show that the CTEM is insensitive to electromagnetic effects~\cite{Holod2013,Candy2005,Pueschel2008,Candy2010}.
Previous theoretical studies of the CTEM have been largely confined to the electrostatic limit~\cite{Adam1976,Catto1978,Tang1978,Cheng1981,ChenH2018,ChenH2019,ChenH2022PRL,Yao2022_1,Yao2022_2}, and the underlying mechanism for this insensitivity remains an open question. 
Therefore, a systematic investigation is required to elucidate the physics governing the weak electromagnetic effects of the CTEM.

In this work, we develop a linear gyrokinetic theory~\cite{ChenL1991,ChenH2024} for the electromagnetic CTEM.
The integro-differential eigenmode equations are derived and solved.
We show that the particle dynamics underlies the weak electromagnetic effects of CTEMs.
Specifically, the trapped-electron parallel current vanishes at leading order because of bounce dynamics.
The subdominant ion transit resonance renders the ion parallel current negligible.
Passing electrons thus dominate the perturbed parallel current in the electromagnetic CTEM regime.
Eigenmode analyses, together with current diagnostics from simulations, validate these findings.

Based on the particle dynamics, a reduced model is constructed for the electromagnetic CTEM by replacing the gyrokinetic vorticity equation with the parallel Amp\`ere's law.
We demonstrate that, in the gyrokinetic vorticity equation, parallel particle dynamics causes the inertia-charge uncovering term \cite{Zonca2006} associated with the magnetic perturbation to vanish.
This eliminates the SAW branch \cite{Chen2021} from the electromagnetic CTEM eigenmode equations.
This decoupling accounts for the weak electromagnetic effects of CTEMs.
We also show that ion parallel dynamics is essential for the coupling between the ITG mode and the SAW branch, and thus for the electromagnetic effects of the ITG modes.

The remainder of this paper is organized as follows.
Section~\ref{sec2} presents the gyrokinetic formalism.
In Sec.~\ref{sec3}, we examine the perturbed parallel current and investigate the electromagnetic coupling between the CTEM and the SAW.
Conclusions are given in Sec.~\ref{sec4}.

\section{Gyrokinetic formalism\label{sec2}}
We investigate the electromagnetic CTEM in a large aspect-ratio ($\epsilon=r/R_0\ll1$), low-$\beta$ ($\beta=8\pi P/B^2\sim\mathcal{O}(\epsilon^2)$) tokamak with shifted circular flux surfaces. 
The $(s, \alpha)$ model~\cite{Connor1978} is adopted, where $s=rq^{\prime}/q$ is the magnetic shear, $q$ is the safety factor, $\alpha=r(\epsilon+\Delta^{\prime})^{\prime}$ is the pressure gradient parameter, and $\Delta$ denotes the Shafranov shift.
With the ordering $\beta\sim\mathcal{O}(\epsilon^2)$, the parameter $\alpha$ can be simplified to $-R_0 q^2 \beta^{\prime}$.
In the toroidal coordinate system $(r, \theta, \zeta)$, where $\theta$ is the poloidal angle and $\zeta$ is the toroidal angle, the background magnetic field takes the form $\boldsymbol{B}=B_0\left[R_0\boldsymbol{\nabla}\zeta + (r/q)\boldsymbol{\nabla}\zeta\times\boldsymbol{\nabla}r\right]$.
Furthermore, the background distribution for species $j$ ($j=i,e$) is taken to be a local Maxwellian, $F_{j}=N_{0}(\pi v_{tj}^2)^{-3/2}\exp\left(-v^2/v_{tj}^2\right)$, where $v_{tj}=\sqrt{2T_j/m_j}$ is the thermal velocity.

Following Ref.~\cite{ChenL1991}, low-frequency electromagnetic fluctuations in low-$\beta$ plasmas can be described by the electrostatic potential $\delta \phi$ and the parallel vector potential $\delta A_{\|}$, while the compressional magnetic perturbation $\delta B_{\|}$ is neglected. 
For convenience, we introduce a scalar induced potential $\delta \psi$ via the relation $-c\nabla_{\|}\delta\psi=\partial_{t}\delta A_{\|}$, where $\nabla_{\|}=\boldsymbol{b}\cdot\boldsymbol{\nabla}$ and $\boldsymbol{b}=\boldsymbol{B}/B$.
The parallel electric field is thus given by $\delta E_{\|}=-\nabla_{\|}(\delta \phi-\delta \psi)$.

Employing the straight-field-line coordinates $(r,\Theta,\zeta)$, where $\Theta=\theta-(\epsilon+\Delta^{\prime})\sin\theta$, we can adopt the ballooning representation~\cite{Connor1979} for fluctuations.
For instance, the electrostatic potential is expressed as
\begin{equation}
\delta \phi=\sum_{n}e^{-i \omega t-i n \zeta} \sum_m e^{i m \Theta} \int_{-\infty}^{+\infty} \mathrm{d} \eta e^{i(n q-m) \eta} \delta \tilde{\phi}_n(\eta).\label{eq:ballooning_representation}
\end{equation}
In terms of the guiding-center velocity variables $(\varepsilon=v^2/2, \mu=v_{\perp}^2/2B)$, the perturbed particle distribution function $\delta \tilde{f}_{j, n}$~\cite{ChenL1991} can be decomposed into its adiabatic and nonadiabatic components:
\begin{equation}
\delta \tilde{f}_{j, n}=-\frac{q_j \delta \tilde{\phi}_n}{T_j} F_{j}+J_{0j}\left[\left(1+\frac{\omega_{* j, n}^t}{\omega}\right) J_{0j}\frac{q_j \delta \tilde{\psi}_n}{T_j}F_j+\delta \tilde{K}_{j, n}\right],\label{eq:perturbed_particle_distribution_function_in_ballooning_space}
\end{equation}
where $J_{nj}=J_n(-k_{\perp}\rho_{j})$ is the Bessel function accounting for the finite Larmor radius (FLR) effects, $\rho_{j}=v_{\perp}/\Omega_{cj}$ is the Larmor radius, $\Omega_{cj}=q_j B/m_j c$ is the cyclotron frequency, $k_{\perp}^2=\left[1+(s \eta-\alpha \sin \eta)^2\right] k_\theta^2$, and $k_\theta=n q / r$.
The kinetic compression $\delta \tilde{K}_{j, n}$ satisfies the linear electromagnetic gyrokinetic equation
\begin{equation}
\left(\omega+i \omega_{tj}\frac{\partial}{\partial \eta}+\omega_{dj, n}\right) \delta \tilde{K}_{j, n}=\left(\omega+\omega_{*j, n}^t\right)\frac{q_j}{T_j}\left(\delta \tilde{S}_{1j, n}+\delta \tilde{S}_{2j, n}\right) F_j.\label{eq:gk_equation_in_ballooning_space}
\end{equation}
In Eq.~\eqref{eq:gk_equation_in_ballooning_space}, the free-energy source terms are given by
\begin{equation}
\delta \tilde{S}_{1j, n}=J_{0 j}\left[\left(\delta \tilde{\phi}_n-\delta \tilde{\psi}_n\right)-\frac{\omega_{d j, n}}{\omega} \delta \tilde{\psi}_n\right],\quad \delta \tilde{S}_{2j, n}=-i \frac{\omega_{tj}}{\omega} \frac{\partial k_{\perp}}{\partial \eta} \rho_j J_{1j} \delta \tilde{\psi}_n.\label{eq:linear_free_energy_sources_in_ballooning_space}
\end{equation}
Here, $\omega_{t j}=\hat{v}_{\|}v_{tj}/qR_0$ is the transit frequency, $\omega_{dj,n}=\omega_{*j,n} \epsilon_{n}(\hat{v}_{\perp}^2+2\hat{v}_{\|}^2)(s \eta \sin \eta+\cos \eta-\alpha \sin^2 \eta)$ is the magnetic drift frequency, $\omega_{*j,n}=c k_{\theta} T_j / q_j B L_{n}$ is the diamagnetic drift frequency, and $\omega_{*j,n}^t=\omega_{*j,n}[1+\eta_{j}(\hat{v}^2-3/2)]$. 
The velocity is normalized to $v_{tj}$. 
The density and temperature gradient scale lengths are defined, respectively, as $L_{n}=-(\partial \ln N_{0}/\partial r)^{-1}$ and $L_{t,j}=-(\partial \ln T_{j}/\partial r)^{-1}$, with $\epsilon_{n}=L_{n}/R_0$ and $\eta_{j}=L_{n}/L_{t,j}$. 
Additionally, the pressure gradient parameter is expressed as $\alpha=\sum_j q^2 \beta_j(1+\eta_j)/\epsilon_{n,j}$, where $\beta_{j}=8\pi N_j T_j/B^2$.

The electrostatic potential $\delta \tilde{\phi}_n$ is governed by the quasineutrality condition,
\begin{equation}
\begin{aligned}
&\left(1+\frac{T_i}{T_e}\right) \tilde{\Phi}_{\|}
+\sum_j \frac{T_i}{T_j}\left[\left(1+\frac{\omega_{*j, n}}{\omega}\right)\left(1-\Gamma_{0j}\right)+\frac{\omega_{*j, n}}{\omega} \eta_j b_j\left(\Gamma_{0j}-\Gamma_{1 j}\right)\right] \tilde{\Psi}\\
&-\frac{1}{N_0} \sum_j \frac{q_j}{q_i}\left\langle J_{0j} \delta \tilde{K}_{j, n}\right\rangle_v=0,\label{eq:quasineutrality_in_ballooning_space}
\end{aligned}
\end{equation}
where $\left\langle\cdots\right\rangle_v$ denotes the integration in velocity space.
$\Gamma_{nj}=I_{n}(b_j)\exp(-b_j)$ represents the FLR effect with $I_{n}$ being the modified Bessel function of the first kind and $b_j=k_{\perp}^2\rho_{tj}^2/2=k_{\perp}^2v_{tj}^2/2\Omega_{cj}^2$.
Here, we adopt the normalizations $\tilde{\Phi}=q_i \delta \tilde{\phi}_n/T_i$ and $\tilde{\Psi}=q_i \delta \tilde{\psi}_{n}/T_i$.
The effective potential $\tilde{\Phi}_{\|}=\tilde{\Phi}-\tilde{\Psi}$, as noted above, is related to the parallel electric field.
Following the conventional approach~\cite{ChenL1991, Zonca2006}, the gyrokinetic vorticity equation for the induced potential $\delta \tilde{\psi}_n$ is derived by taking the velocity space moment of the gyrokinetic equation~\eqref{eq:gk_equation_in_ballooning_space}:
\begin{equation}
\begin{aligned}
&\frac{i}{N_0}\sum_j\frac{q_j}{q_i}\left\langle  J_{0 j} \frac{\omega_{t j}}{\omega}\frac{\partial}{\partial \eta} \delta \tilde{K}_{j, n}\right\rangle_v+\frac{1}{N_0}\sum_j \frac{q_j}{q_i}\left\langle J_{0 j} \delta \tilde{K}_{j, n}\right\rangle_v\\
&+\frac{1}{N_0}\sum_j \frac{q_j}{q_i}\left\langle J_{0 j} \frac{\omega_{d j, n}}{\omega} \delta \tilde{K}_{j, n}\right\rangle_v-\frac{1}{N_0}\sum_j\frac{T_i}{T_j}\left\langle\left(1+\frac{\omega_{*j, n}^t}{\omega}\right) J_{0 j}^2 \tilde{\Phi}_{\|} F_j\right\rangle_v\\
&+\frac{1}{N_0}\sum_j \frac{T_i}{T_j}\left\langle\left(1+\frac{\omega_{*j, n}^t}{\omega}\right) J_{0 j}^2 \frac{\omega_{d j, n}}{\omega} \tilde{\Psi} F_j\right\rangle_v=0.\label{eq:moment_of_gk}
\end{aligned}
\end{equation}
Using the parallel Amp\`ere's law with the Coulomb gauge, the first term is identified as the field line bending (FLB) term,
\begin{equation}
\text{FLB}=\frac{\omega_A^2}{\omega^2} \frac{\partial}{\partial \eta}\left(b_i \frac{\partial \tilde{\Psi}}{\partial \eta}\right)-\frac{i}{N_0} \sum_j\frac{q_j}{q_i}\left\langle J_{1 j} \frac{\partial k_{\perp}}{\partial \eta} \rho_j \frac{\omega_{t j}}{\omega} \delta \tilde{K}_{j, n}\right\rangle _v,
\end{equation}
where $\omega_A^2=v_A^2/q^2 R_0^2$ is the Alfv\'en frequency and $v_A=B/\sqrt{4\pi N_0 m_i}$.
The second and fourth terms of Eq.~\eqref{eq:moment_of_gk} combine to give the inertia-charge uncovering (ICU) term \cite{Zonca2006},
\begin{equation}
\text{ICU}=\frac{1}{N_0}\sum_j \frac{q_j}{q_i}\left\langle J_{0 j} \delta \tilde{K}_{j, n}\right\rangle_v-\frac{1}{N_0}\sum_j\frac{T_i}{T_j}\left\langle\left(1+\frac{\omega_{*j, n}^t}{\omega}\right) J_{0 j}^2 \tilde{\Phi}_{\|} F_j\right\rangle_v.
\end{equation}
Here, the nonadiabatic charge separation term $(1/N_0)\sum_j(q_j/q_i)\langle J_{0 j} \delta \tilde{K}_{j, n}\rangle_v$ is denoted as $\text{ICU}_1$, and the parallel electric field term $-(1/N_0)\sum_j(T_i/T_j)\langle(1+(\omega_{*j, n}^t/\omega)) J_{0 j}^2 \tilde{\Phi}_{\|} F_j\rangle_v$ is denoted as $\text{ICU}_2$.
Utilizing the quasineutrality condition~\eqref{eq:quasineutrality_in_ballooning_space}, the $\text{ICU}$ term is then expressed as $\text{ICU}=\text{ICU}_{\tilde{\Phi}_{\|}}+\text{ICU}_{\tilde{\Psi}}$, with $\text{ICU}_{\{\tilde{\Phi}_{\|},\tilde{\Psi}\}}=\sum_j (T_i/T_j)[(1+\omega_{*j, n}/\omega)(1-\Gamma_{0 j})+(\omega_{*j, n}/\omega)\eta_j b_j(\Gamma_{0 j}-\Gamma_{1 j})]\{\tilde{\Phi}_{\|},\tilde{\Psi}\}$.
The third and fifth terms of Eq.~\eqref{eq:moment_of_gk} represent, respectively, the kinetic nonadiabatic particle compression (KPC) and the magnetohydrodynamic (MHD) nonadiabatic particle compression (MPC):
\begin{equation}
\begin{aligned}
\text{KPC}&=\frac{1}{N_0}\sum_j \frac{q_j}{q_i}\left\langle J_{0 j} \frac{\omega_{d j, n}}{\omega} \delta \tilde{K}_{j, n}\right\rangle_v,\\
\text{MPC}&=\frac{1}{N_0}\sum_j \frac{T_i}{T_j}\left\langle\left(1+\frac{\omega_{*j, n}^t}{\omega}\right) J_{0 j}^2 \frac{\omega_{d j, n}}{\omega} \tilde{\Psi} F_j\right\rangle_v.\label{eq:kpc_mpc_vorticity_equation}
\end{aligned}
\end{equation}
Consequently, the gyrokinetic vorticity equation takes the form
\begin{equation}
\text{FLB}+\text{ICU}+\text{KPC}+\text{MPC}=0.\label{eq:vorticity_equation_in_ballooning_space}
\end{equation}
In the shearless uniform slab geometry and ideal MHD limit ($\tilde{\Phi}_{\|}=0$), the balance between the $\text{FLB}$ term and the $\text{ICU}_{\tilde{\Psi}}$ term recovers the SAW dispersion relation \cite{Chen2021}.
Therefore, Eqs.~\eqref{eq:quasineutrality_in_ballooning_space} and \eqref{eq:vorticity_equation_in_ballooning_space} describe the electromagnetic coupling between the drift wave and the SAW branch.

For the electromagnetic CTEM under the typical drift ordering, the conditions $\epsilon_{n}\lesssim\mathcal{O}(1)$ and $k_{\theta}\rho_{ti}\sim\mathcal{O}(1)$ indicate
\begin{equation}
\omega_{ti} < \omega_{di,n} \sim \omega_{de,n} \sim \left\langle\omega_{de,n}\right\rangle_T \sim \omega \ll \omega_{be}<\omega_{te},\label{eq:ordering}
\end{equation}
where $\omega_{be}\sim\sqrt{\epsilon} \omega_{te}$ is the trapped-electron bounce frequency and $\langle\omega_{de,n}\rangle_T$ is the trapped-electron precession frequency.
Based on the frequency ordering \eqref{eq:ordering}, we expand the ion gyrokinetic equation in powers of $\omega_{t i} / \omega_{di,n}$ and obtain the leading-order kinetic compression:
\begin{equation}
\delta \tilde{K}_{i, n}^{(0)}=\frac{\omega+\omega_{* i, n}^t}{\omega+\omega_{d i, n}}J_{0i}\left( \tilde{\Phi}_{\|}-\frac{\omega_{d i, n}}{\omega} \tilde{\Psi}\right) F_{i}.\label{eq:ion_zero_order_in_ballooning_space}
\end{equation}
In this short-wavelength regime, the ion kinetic effect is dominated by the ``drift resonance'', while the transit resonance is negligible \cite{Romanelli1990, Romanelli1989, Coppi1974}.
For trapped electrons, the leading-order kinetic compression can be derived from the bounce-kinetic equation~\cite{Tang1978,Cheng1981,ChenH2018,ChenH2019,ChenH2022PRL}:
\begin{equation}
\begin{aligned}
\delta \tilde{K}_{t e, n}^{(0)}=-\frac{T_i}{T_e}\frac{\omega+\omega_{* e, n}^t}{\omega+\left\langle\omega_{d e, n}\right\rangle_T}F_{e}\sum_{l=-\infty}^{\infty} \int_{2 l \pi-\eta_0}^{2 l \pi+\eta_0} \mathrm{d} \eta^{\prime} \delta\left(\eta-\eta^{\prime}\right)\left\langle\tilde{\Phi}_{\|}-\frac{\omega_{d e, n}}{\omega} \tilde{\Psi}\right\rangle_{T}.\label{eq:trapped_electron_lowest_order}
\end{aligned}
\end{equation}
The bounce-average operator is defined as
\begin{equation}
\left\langle\tilde{\Phi}_{\|}\right\rangle_{T}=\frac{1}{4 K(\kappa)} \int_{2 l \pi-\eta_0}^{2 l \pi+\eta_0} \mathrm{d} \vartheta \frac{\tilde{\Phi}_{\|}(\vartheta)}{\sqrt{\kappa^2-\sin^2(\vartheta/2)}} .\label{eq:bounce_average}
\end{equation}
The precession frequency of trapped electrons is given by $\left\langle\omega_{de,n}\right\rangle_T=\omega_{*e,n} \epsilon_{n, e} \hat{v}^2 H$, where $H$ is defined as
\begin{equation}
H=4 s\left[\frac{E(\kappa)}{K(\kappa)}+\kappa^2-1\right]+2 \frac{E(\kappa)}{K(\kappa)}-1-\frac{4}{3} \alpha\left[\left(2 \kappa^2-1\right) \frac{E(\kappa)}{K(\kappa)}-\left(\kappa^2-1\right)\right].\label{eq:electron_precessional_frequency}
\end{equation}
Here, $K(\kappa)$ and $E(\kappa)$ are the complete elliptic integrals of the first and second kind, respectively.
The pitch-angle variable is defined as $\kappa^2=[\varepsilon-\mu B_0 (1-\epsilon)]/2\epsilon \mu B_0$, and $\eta_0 = 2\arcsin\kappa$ represents the turning point.

Substituting Eqs.~\eqref{eq:ion_zero_order_in_ballooning_space}--\eqref{eq:electron_precessional_frequency} into Eqs.~\eqref{eq:quasineutrality_in_ballooning_space} and \eqref{eq:vorticity_equation_in_ballooning_space} yields the eigenmode equations for the electromagnetic CTEM:
\begin{equation}
\begin{aligned}
&Q_1 \tilde{\Phi}_{\|}+\left(V_1+V_2\right) \tilde{\Psi}-\tau\sqrt{\frac{2\epsilon}{\pi}}\int_{\sin^2(\eta/2)}^1 \frac{\mathrm{d} \kappa^2}{\sqrt{\kappa^2-\sin^2(\eta/2)}} \\
\times&\sum_{l=-\infty}^{\infty} \int_{2l\pi-\pi}^{2 l\pi+\pi}\mathrm{d} \eta^{\prime}\delta\left(\eta-\eta^{\prime}\right)\left\langle\tilde{\Phi}_{\|}-\frac{\hat{\omega}_{d e, n}}{\hat{\omega}} \tilde{\Psi}\right\rangle_{T}\int_0^{\infty}\mathrm{d} \hat{v}^2 \frac{\hat{\omega}+\hat{\omega}_{*e,n}^t}{\hat{\omega}+\left\langle\hat{\omega}_{de,n}\right\rangle_T}\hat{v} e^{-\hat{v}^2}=0,\label{eq:eigenmode_quasineutrality}\\
\end{aligned}
\end{equation}
\begin{equation}
\begin{aligned}
&\left(V_1+V_2\right)\tilde{\Phi}_{\|}+\left(V_1+V_2+V_3+V_4\right) \tilde{\Psi}+\tau\sqrt{\frac{2\epsilon}{\pi}}\int_{\sin^2(\eta/2)}^1 \frac{\mathrm{d} \kappa^2}{\sqrt{\kappa^2-\sin^2(\eta/2)}} \\
\times&\sum_{l=-\infty}^{\infty} \int_{2l\pi-\pi}^{2l\pi+\pi}\mathrm{d} \eta^{\prime}\delta\left(\eta-\eta^{\prime}\right)\left\langle\tilde{\Phi}_{\|}-\frac{\hat{\omega}_{d e, n}}{\hat{\omega}} \tilde{\Psi}\right\rangle_{T}\int_0^{\infty}\mathrm{d} \hat{v}^2\frac{\hat{\omega}_{de,n}}{\hat{\omega}}\frac{\hat{\omega}+\hat{\omega}_{*e,n}^t}{\hat{\omega}+\left\langle\hat{\omega}_{de,n}\right\rangle_T}\hat{v}e^{-\hat{v}^2}=0, \label{eq:eigenmode_vorticity}
\end{aligned}
\end{equation}
where the coefficients are given by
\begin{equation}
\begin{aligned}
Q_1=&\left(1+\tau\right)-\left\langle\frac{\hat{\omega}+\hat{\omega}_{*i, n}^t}{\hat{\omega}+\hat{\omega}_{d i, n}} J_{0 i}^2 \frac{F_{i}}{N_0}\right\rangle_v, \quad \tau= \frac{T_i}{T_e},\\
V_1=&\left(1+\frac{1}{\hat{\omega}}\right)\left(1-\Gamma_{0 i}\right)+\frac{1}{\hat{\omega}} \eta_i b_i\left(\Gamma_{0 i}-\Gamma_{1 i}\right),\quad V_2=\left\langle\frac{\hat{\omega}+\hat{\omega}_{*i, n}^t}{\hat{\omega}+\hat{\omega}_{d i, n}} \frac{\hat{\omega}_{d i, n}}{\hat{\omega}} J_{0 i}^2 \frac{F_{i}}{N_0}\right\rangle_v,\\
V_3=&\frac{2 \epsilon_{n}^2}{\tau \beta_e q^2 \hat{\omega}^2 k_\theta^2} \frac{\partial}{\partial \eta}\left(k_{\perp}^2 \frac{\partial}{\partial \eta}\right), \quad V_4=\tau\left\langle\left(1+\frac{\hat{\omega}_{* e, n}^t}{\hat{\omega}}\right)\frac{\hat{\omega}_{d e, n}}{\hat{\omega}} \frac{F_e}{N_0}\right\rangle_v.\label{eq:eigenmode_coefficients}
\end{aligned}
\end{equation}
In Eqs.~\eqref{eq:eigenmode_quasineutrality}--\eqref{eq:eigenmode_coefficients}, the frequencies have been normalized to the ion diamagnetic drift frequency $\omega_{*i,n}$. 
Additionally, in the ion velocity space integration, the $\boldsymbol{\nabla} B$ drift contribution to $\omega_{di,n}$ has been neglected~\cite{Romanelli1989}.
Comparing equation~\eqref{eq:vorticity_equation_in_ballooning_space} with the equation~\eqref{eq:eigenmode_vorticity}, we identify $V_1\tilde{\Psi}$ as the $\text{ICU}_{\tilde{\Psi}}$ term and $V_3\tilde{\Psi}$ as the $\text{FLB}$ term.
Thus, the eigenmode equations~\eqref{eq:eigenmode_quasineutrality} and \eqref{eq:eigenmode_vorticity} describe the electromagnetic coupling between the CTEM and the SAW branch.
As will be shown in Sec.~\ref{sec3.5}, the particle dynamics of ions and trapped electrons decouple the CTEM from the SAW branch.

\section{Results\label{sec3}}
\subsection{Electromagnetic effects\label{sec3.1}}
\begin{figure}[htbp]
\centering
\includegraphics{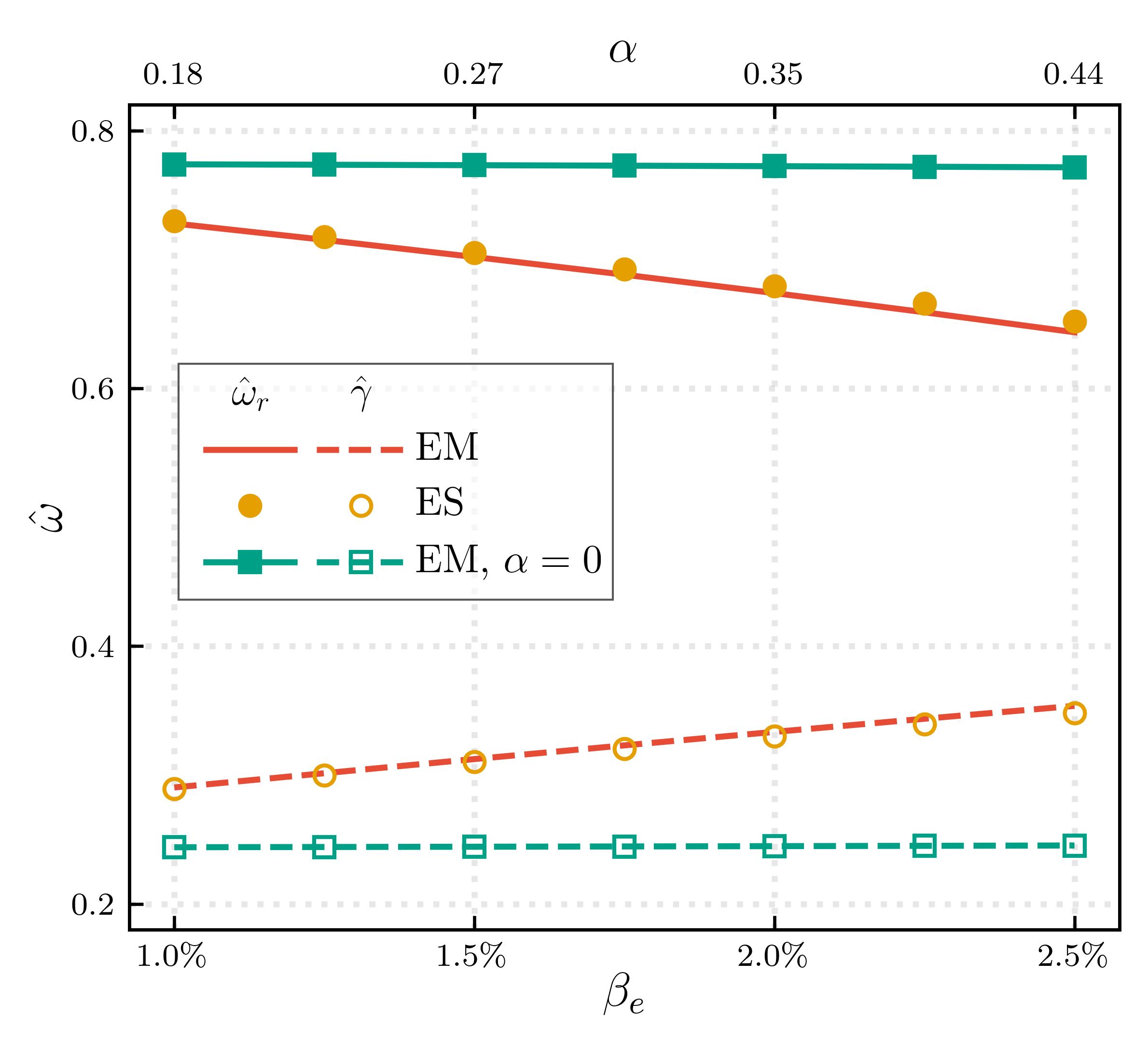}
\caption{\label{fig1:CTEM_eigenfrequencies}CTEM eigenfrequencies versus $\beta_{e}$ for the full electromagnetic (EM), electrostatic (ES), and $\alpha=0$ EM models. The parameters are $\epsilon=0.18$, $k_{\theta}\rho_{ti}=1.33$, $q=1.41$, $s=0.83$, $\epsilon_{n}=0.45$, $\eta_{i}=0$, $\eta_{e}=2.0$, and $\tau=1$.}
\end{figure}

\begin{figure}[htbp]
\centering
\includegraphics{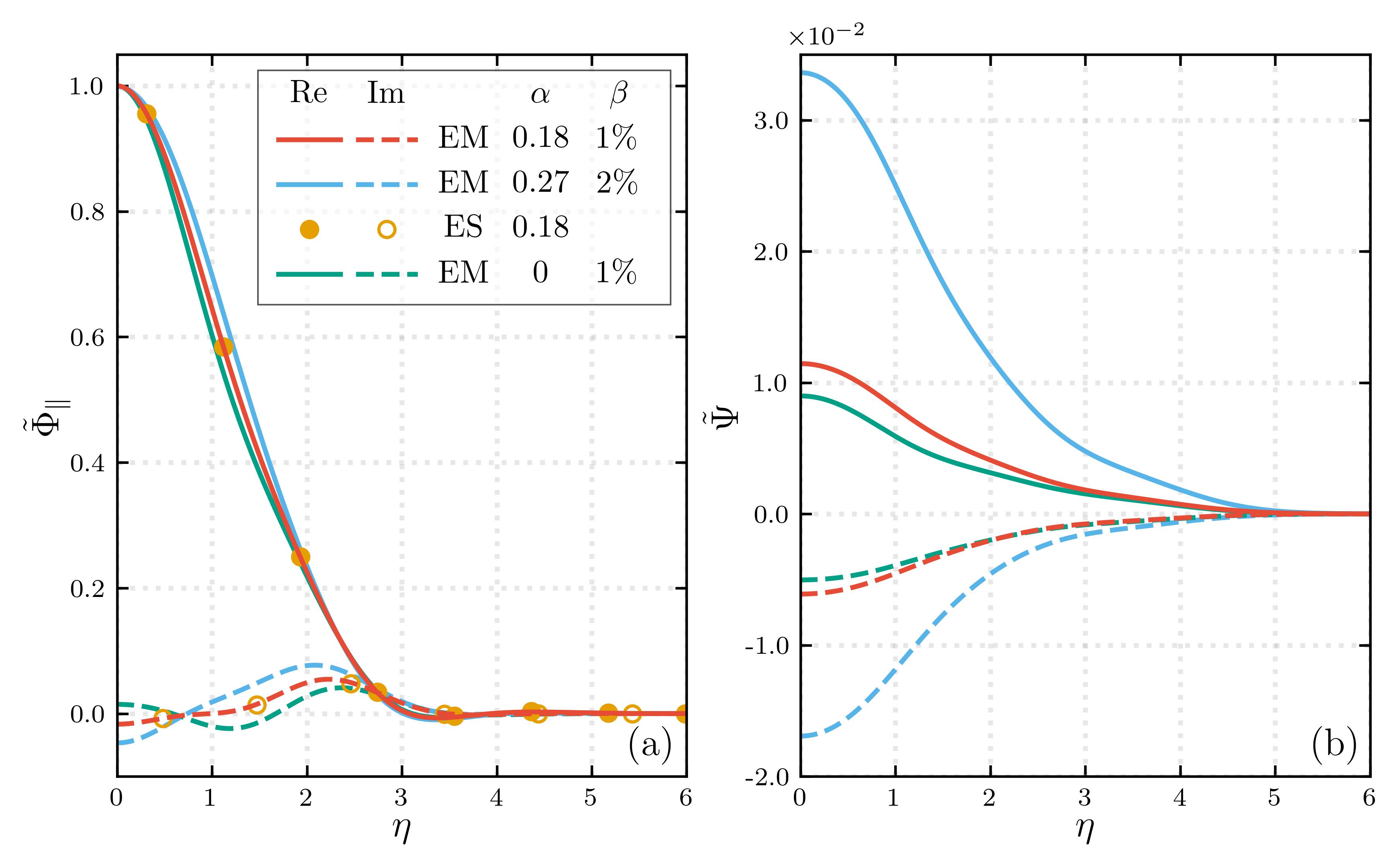}
\caption{\label{fig2:CTEM_eigenmode_structures}CTEM eigenmode structures of (a) $\tilde{\Phi}_{\|}$ and (b) $\tilde{\Psi}$ for the full EM, ES, and $\alpha=0$ EM models.
The parameters are the same as in Fig.~\ref{fig1:CTEM_eigenfrequencies}.}
\end{figure}

The integro-differential eigenmode equations~\eqref{eq:eigenmode_quasineutrality} and \eqref{eq:eigenmode_vorticity} are solved subject to outgoing-wave boundary conditions.
We apply the Galerkin method with a Hermite-Gaussian basis, transforming the integro-differential eigenvalue problem into a generalized matrix eigenvalue problem.
Physical solutions correspond to the vanishing determinant of this matrix.
The roots of the determinant are then located using the Zero Pole Location code~\cite{ChenH2022,Li2025}.
Figure~\ref{fig1:CTEM_eigenfrequencies} illustrates the dependence of the electromagnetic (EM) CTEM eigenfrequency on $\beta_e$.
The corresponding eigenmode structures are given in Fig.~\ref{fig2:CTEM_eigenmode_structures}.
Due to the even parity of the EM CTEM in ballooning space, only $\eta\geq0$ is plotted.
Note that plasma $\beta$ plays a dual role: it determines the pressure gradient parameter $\alpha$ through the Shafranov shift, and it modifies the $\text{FLB}$ term in the gyrokinetic vorticity equation~\eqref{eq:eigenmode_vorticity}.
These mechanisms are referred to, respectively, as the $\alpha$ effect and the magnetic perturbation effect.
To isolate these effects, we compare the full EM model with two reduced models: (i) an electrostatic (ES) model with finite $\alpha$, and (ii) an EM model with $\alpha=0$.

Figure~\ref{fig1:CTEM_eigenfrequencies} shows that the $\alpha=0$ EM CTEM is insensitive to $\beta_e$, indicating negligible magnetic perturbation effects.
Crucially, the agreement between the full EM and ES models indicates that the CTEM is decoupled from the SAW branch, as will be demonstrated in Sec.~\ref{sec3.5}.
According to the gyrokinetic vorticity equation~\eqref{eq:eigenmode_vorticity}, the magnetic perturbation scales as $\tilde{\Psi} / \tilde{\Phi}_{\|} \propto \mathcal{O}\left(\beta_e\right)$.
Consistent with this scaling, Fig.~\ref{fig2:CTEM_eigenmode_structures} shows that the amplitude of $\tilde{\Psi}$ increases with $\beta_e$, yet its influence on CTEM stability remains subdominant.
Furthermore, the eigenmode structures of the EM CTEM extend from $-\pi$ to $\pi$ in ballooning space, comparable to those of the EM ITG mode~\cite{Dong1999}.
However, unlike the ITG mode~\cite{Kim1993}, the CTEM is insensitive to electromagnetic effects. 
This contrast indicates that the different electromagnetic properties of these instabilities cannot be attributed to their eigenmode structures, thereby motivating a detailed investigation of the perturbed parallel current.

\subsection{Parallel current analysis\label{sec3.2}}
We systematically analyze the perturbed parallel current, $\delta \tilde{J}_{\|}=\sum_j \delta \tilde{J}_{j \|}$, where $\delta \tilde{J}_{j \|}= q_j\langle v_{\|} \delta \tilde{f}_{j, n}\rangle_v= q_j\langle v_{\|} J_{0 j} \delta \tilde{K}_{j, n}\rangle_v$.
For ions, the subdominant transit resonance renders the leading-order kinetic compression $\delta \tilde{K}_{i,n}^{(0)}$ an even function of $\hat{v}_{\|}$.
Consequently, the perturbed ion parallel current $\delta \tilde{J}_{i\|}$ is negligible in the CTEM regime, as will be shown in Fig.~\ref{fig5:CTEM_GEM_currents_1} (a).

By taking the velocity space moment of the gyrokinetic equation, the electron parallel current can be decomposed into two components:
\begin{equation}
B \nabla_{\|}\left(\frac{1}{B}\delta \tilde{J}_{e \|}\right)=\tilde{R}_{k, e}+\tilde{R}_{f, e},
\end{equation}
where the kinetic component, $\tilde{R}_{k, e}=i q_e\langle(\omega+\omega_{de,n}) \delta \tilde{K}_{e, n}\rangle_{v}$, arises from the compression term $\delta \tilde{K}_{e, n}$, and the fluid-like component, $\tilde{R}_{f, e}=i q_e\langle(\omega+\omega_{* e, n}^t)[\tilde{\Phi}_{\|}-(\omega_{d e, n} / \omega) \tilde{\Psi}] \tau F_e\rangle_v$, stems from the free-energy source term $\delta \tilde{S}_{1e,n}$.
These components are evaluated in two steps.
First, we consider the passing- and trapped-electron fluid-like components, denoted by $\tilde{R}_{f, pe}$ and $\tilde{R}_{f,te}$, respectively. 
Given that these two terms scale with the population fraction, they are of comparable magnitude in modern tokamaks, $\tilde{R}_{f,te} / \tilde{R}_{f, pe} \sim\sqrt{2 \epsilon} /(1-\sqrt{2 \epsilon}) \sim \mathcal{O}(1)$.
Second, we examine the kinetic components.
With the frequency ordering $\omega \sim\langle\omega_{d e, n}\rangle_T \ll \omega_{t e}$, the trapped-electron kinetic compression, $\delta\tilde{K}_{te, n}$, is governed by the toroidal precession resonance, whereas passing electrons remain adiabatic.
Hence, the passing-electron kinetic component is negligible compared to that of trapped electrons, $\tilde{R}_{k, pe} / \tilde{R}_{k, te}\sim \mathcal{O}(\omega/\omega_{te})\ll \mathcal{O}(1)$.

Furthermore, a fundamental cancellation mechanism governs the trapped-electron parallel current.
The bounce-kinetic equation shows that the leading-order kinetic compression of trapped electrons is constant along the unperturbed orbits, $\partial_{\eta}\delta \tilde{K}_{te,n}^{(0)}=0$.
The bounce dynamics leads to the cancellation between the kinetic and fluid-like components of the trapped-electron parallel current:
\begin{equation}
\tilde{R}_{k, te}+\tilde{R}_{f,te}\simeq q_e\left\langle\omega_{t e}\frac{\partial}{\partial \eta} \delta \tilde{K}_{te, n}^{(0)}\right\rangle_{v,te}=0.\label{eq:te_cancellation}
\end{equation}
Thus, the net trapped-electron parallel current vanishes at leading order.

In summary, combining Eq.~\eqref{eq:te_cancellation} with the orderings $\tilde{R}_{f, p e} / \tilde{R}_{k, t e} \sim \mathcal{O}(1)$ and $\tilde{R}_{k, p e} / \tilde{R}_{k, t e} \ll \mathcal{O}(1)$ reveals that the perturbed parallel current in the electromagnetic CTEM is dominated by the passing-electron fluid-like contribution, $\tilde{R}_{f,pe}$.
Since $\tilde{R}_{f,pe}\propto \tilde{\Phi}_{\|}$, this current is expected to oscillate in phase with the electrostatic potential, as will be shown in Fig.~\ref{fig6:CTEM_GEM_currents_2}.
It is worth emphasizing that the disparity in current contributions stems from the distinct particle dynamics of passing and trapped electrons.
Passing electrons stream freely along unperturbed orbits, whereas the parallel velocity of trapped electrons averages to zero over a bounce period.

\begin{figure}[htbp]
\centering
\includegraphics{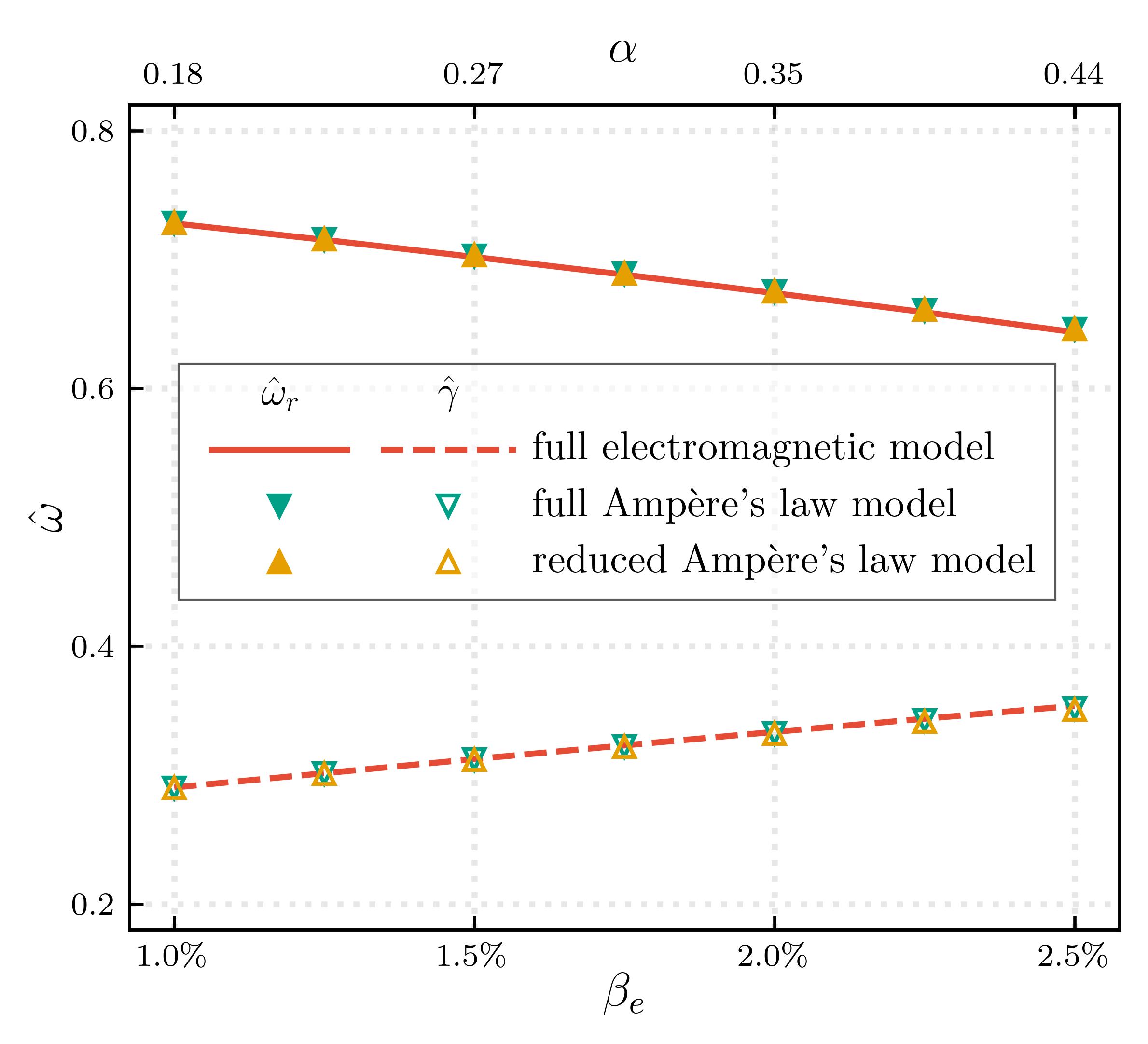}
\caption{\label{fig3:CTEM_simplified_model_eigenfrequencies}CTEM eigenfrequencies versus $\beta_e$ for the full electromagnetic model (Eqs.~\eqref{eq:eigenmode_quasineutrality} and \eqref{eq:eigenmode_vorticity}), the full Amp\`ere's law model (Eqs.~\eqref{eq:eigenmode_quasineutrality} and \eqref{eq:simplified_model}), and the reduced Amp\`ere's law model (Eqs.~\eqref{eq:eigenmode_quasineutrality} and \eqref{eq:simplified_model}, retaining only the passing-electron fluid-like contribution in the latter).
The parameters are the same as in Fig.~\ref{fig1:CTEM_eigenfrequencies}.}
\end{figure}
\begin{figure}[htbp]
\centering
\includegraphics[width=\textwidth]{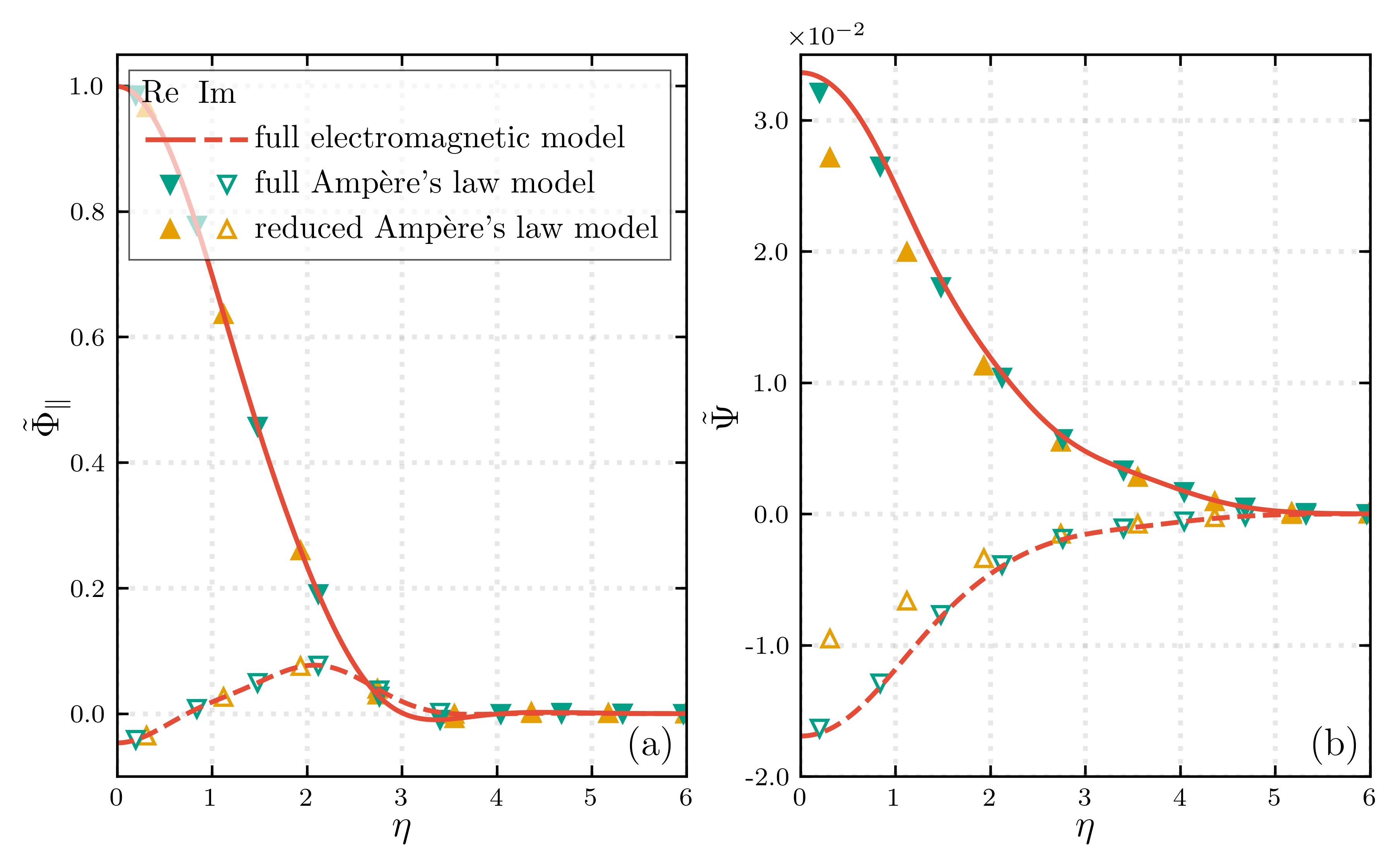}
\caption{\label{fig4:CTEM_simplified_model_modestructures}CTEM eigenmode structures of (a) $\tilde{\Phi}_{\|}$ and (b) $\tilde{\Psi}$ for the full electromagnetic, full Amp\`ere's law, and reduced Amp\`ere's law models.
The parameters are the same as in Fig.~\ref{fig1:CTEM_eigenfrequencies}, except for $\beta_e=1\%$.}
\end{figure}

To corroborate these analytical findings, we construct a reduced electromagnetic CTEM model by replacing the gyrokinetic vorticity equation \eqref{eq:eigenmode_vorticity} with the parallel Amp\`ere's law, $c^2 \nabla_{\perp}^2 \nabla_{\|} \partial_t^{-1} \delta \psi=4 \pi (\delta J_{pe\|} + \delta J_{te\|})$.
The quasineutrality condition, Eq.~\eqref{eq:eigenmode_quasineutrality}, remains unchanged.
The governing equation for the induced potential is expressed as
\begin{equation}
V_3 \tilde{\Psi}+\hat{\tilde{R}}_{f, p e}+\hat{\tilde{R}}_{f, t e}+\hat{\tilde{R}}_{k, t e}=0,\label{eq:simplified_model}
\end{equation}
where the normalized source terms are given by
\begin{equation}
\begin{aligned}
\hat{\tilde{R}}_{f, p e}=&\left[1-\sqrt{2 \epsilon\left(1-\sin^2 \frac{\eta}{2}\right)}\right] \hat{\tilde{R}}_{f,e},\quad \hat{\tilde{R}}_{f, t e}=\sqrt{2 \epsilon\left(1-\sin^2 \frac{\eta}{2}\right)} \hat{\tilde{R}}_{f,e},\\
\hat{\tilde{R}}_{f,e}=&-\tau\left\langle\left(1+\frac{\hat{\omega}_{*e, n}^t}{\hat{\omega}}\right) \frac{F_e}{N_0}\right\rangle_v \tilde{\Phi}_{\|}+\tau\left\langle\left(1+\frac{\hat{\omega}_{* e, n}^t}{\hat{\omega}}\right) \frac{\hat{\omega}_{d e, n}}{\hat{\omega}} \frac{F_e}{N_0}\right\rangle_v \tilde{\Psi},\\
\hat{\tilde{R}}_{k,te}=&\tau\sqrt{\frac{2 \epsilon}{\pi}}\int_{\sin^2 (\eta/2)}^1 \frac{\mathrm{d} \kappa^2}{\sqrt{\kappa^2-\sin^2(\eta/2)}}\sum_{l=-\infty}^{\infty} \int_{2l\pi-\pi}^{2l\pi+\pi}\mathrm{d} \eta^{\prime}\delta\left(\eta-\eta^{\prime}\right) \\
&\times\left\langle\tilde{\Phi}_{\|}-\frac{\hat{\omega}_{d e, n}}{\hat{\omega}} \tilde{\Psi}\right\rangle_{T}\int_0^{\infty}\mathrm{d} \hat{v}^2\left(1+\frac{\hat{\omega}_{de,n}}{\hat{\omega}}\right)\frac{\hat{\omega}+\hat{\omega}_{*e,n}^t}{\hat{\omega}+\left\langle\hat{\omega}_{de,n}\right\rangle_T}\hat{v}e^{-\hat{v}^2}.
\end{aligned}
\end{equation}

Guided by the preceding particle dynamics analysis, we evaluate the specific currents by testing two variants of this reduced model.
First, we consider a reduced Amp\`ere's law model that retains only the passing-electron fluid-like contribution $\hat{\tilde{R}}_{f, p e}$.
This reduced model is expected to reproduce the eigenfrequencies and eigenmode structures obtained from the full gyrokinetic formulation (Eqs.~\eqref{eq:eigenmode_quasineutrality} and \eqref{eq:eigenmode_vorticity}, termed the full electromagnetic model in Sec.~\ref{sec3.1}).
Second, we consider a full Amp\`ere's law model that retains all source terms, $\hat{\tilde{R}}_{f, p e}$ and $\hat{\tilde{R}}_{f, t e}+\hat{\tilde{R}}_{k, t e}$, which enables a numerical verification of the cancellation mechanism of the trapped-electron current.
Comparisons between the reduced Amp\`ere's law model, the full Amp\`ere's law model, and the full electromagnetic model are presented in Figs.~\ref{fig3:CTEM_simplified_model_eigenfrequencies} and \ref{fig4:CTEM_simplified_model_modestructures}.
The reduced Amp\`ere's law model is in agreement with the full electromagnetic model, validating that the perturbed parallel current is dominated by the fluid-like contribution of passing electrons.
Furthermore, retaining both kinetic and fluid-like trapped-electron currents in the full Amp\`ere's law model gives identical eigenfrequencies and introduces only higher-order corrections to the eigenmode structures.
This confirms the cancellation mechanism described by Eq.~\eqref{eq:te_cancellation}.
As will be analyzed in the following subsection, the higher-order residual trapped-electron current responsible for these subdominant corrections arises from the nonzero bounce harmonics of the electrostatic potential.

\subsection{Trapped-electron current contribution\label{sec3.3}}
We explicitly evaluate the trapped-electron parallel current by solving the gyrokinetic equation using a Fourier transform method~\cite{ChenH2025}.
To exploit the periodicity of the bounce motion, we introduce the angle variable $\eta_{b}=\omega_{be}\int^{\eta}\mathrm{d}\eta^{\prime}/{\dot{\eta}^{\prime}}$, which is canonically conjugate to the second invariant~\cite{ChenL2016}.
Here, $\dot{\eta}^{\prime}=v_{\|}/{q R_0}$, and $T=\oint \mathrm{d} \eta / \dot{\eta}=2 \pi / \omega_{b e}$ is the bounce period.
We define the Fourier transform as
\begin{equation}
\mathcal{F}_b[g]_{\nu}=\frac{1}{T} \oint \frac{\mathrm{d} \eta}{\dot{\eta}} g(\eta) e^{-i \nu \omega_{b e} \int^\eta \mathrm{d} \eta^{\prime}/\dot{\eta}^{\prime}}.
\end{equation}
Consequently, a perturbed quantity $g$ can be decomposed into its bounce harmonics:
\begin{equation}
g=\sum_{\nu=-\infty}^{\infty}  e^{i\nu \omega_{b e} \int^\eta \mathrm{d} \eta^{\prime}/\dot{\eta}^{\prime}} \mathcal{F}_b[g]_\nu.
\end{equation}
Specifically, the $\nu=0$ harmonic corresponds to the bounce-averaged result, i.e., $\mathcal{F}_b[g]_0=\langle g\rangle_{T}$.
The gyrokinetic equation of trapped electrons then takes the form:
\begin{equation}
\left(\omega+i \omega_{b e}\frac{\partial}{\partial \eta_{b}}+\omega_{d e, n}\right) \delta \tilde{K}_{t e, n}=-\left(\omega+\omega_{*e, n}^t\right)\left(\tilde{\Phi}_{\|}-\frac{\omega_{d e, n}}{\omega} \tilde{\Psi}\right)\tau F_{e}.\label{eq:trapped_electron_gk_in_eta_b}
\end{equation}
The magnetic drift frequency $\omega_{de,n}$ can be Fourier decomposed as $\omega_{d e, n}\simeq\langle\omega_{d e, n}\rangle_T+\omega_{d e, n}^{(1)} \cos \eta_b+\omega_{d e, n}^{(2)} \cos 2 \eta_b$, where $\omega_{d e, n}^{(1,2)}=\mathcal{F}_b\left[\omega_{d e, n}\right]_{1,2}$.
Transforming Eq.~\eqref{eq:trapped_electron_gk_in_eta_b} to the banana center using the Euler factor $\exp (-i L_b)=\exp[-i\sum_{\sigma=1,2}(1/\sigma)(\omega_{d e, n}^{(\sigma)}/\omega_{b e}) \sin(\sigma\eta_b)]$ yields
\begin{equation}
\begin{aligned}
&\left(\omega+\left\langle\omega_{d e, n}\right\rangle_T+i \omega_{b e} \frac{\partial}{\partial \eta_{b}}\right)\left(e^{-i L_b}\delta \tilde{K}_{t e, n}\right)=-\tau F_e e^{-i L_b}\left(\omega+\omega_{* e, n}^t\right) \\
&\times\sum_{\nu=-\infty}^\infty\left\{\mathcal{F}_b\left[\tilde{\Phi}_{\|}-\frac{\left\langle\omega_{d e, n}\right\rangle_T}{\omega} \tilde{\Psi}\right]_\nu-\sum_{\sigma=1,2} \frac{\omega_{d e, n}^{(\sigma)} \cos \sigma \eta_b}{\omega} \mathcal{F}_b\left[\tilde{\Psi}\right]_\nu\right\} e^{i \nu \eta_b}.\label{eq:trapped_electron_gk_in_banana_center}
\end{aligned}
\end{equation}
The solution of Eq.~\eqref{eq:trapped_electron_gk_in_banana_center} is readily obtained as
\begin{equation}
\begin{aligned}
\delta \tilde{K}_{te, n}=&-e^{i L_b} \sum_{p_1, p_2, \nu}J_{p_1} J_{p_2}\frac{\omega+\omega_{* e, n}^t}{\Omega_{p_1, p_2, \nu}}\mathcal{F}_b\left[\tilde{\Phi}_{\|}-\frac{\left\langle\omega_{d e, n}\right\rangle_T}{\omega} \tilde{\Psi}\right]_\nu e^{-i\left(p_1+2 p_2-\nu\right) \eta_b} \tau F_e\\
&-e^{i L_b} \sum_{p_1, p_2, \nu} \sum_{\sigma= \pm 1, \pm 2} J_{p_1} J_{p_2}\frac{\omega_{d e, n}^{(|\sigma|)}}{2\omega}\frac{\omega+\omega_{* e, n}^t}{\Omega_{p_1,p_2, \nu-\sigma}} \mathcal{F}_b\left[\tilde{\Psi}\right]_\nu e^{-i\left(p_1+2 p_2-\nu+\sigma\right) \eta_b} \tau F_e,\label{eq:solution}
\end{aligned}
\end{equation}
where $\Omega_{p_1, p_2, \nu}=\omega+\langle\omega_{d e, n}\rangle_T+(p_1+2 p_2-\nu) \omega_{b e}$.
The Bessel functions $J_{p_1}=J_{p_1}(\omega_{d e, n}^{(1)} / \omega_{b e})$ and $J_{p_2}=J_{p_2}(\omega_{d e, n}^{(2)} / 2 \omega_{b e})$ account for the finite banana orbit width effects.
With the orderings $|\omega_{d e, n}^{(1,2)}/\omega_{b e}|\ll 1$ and $|\tilde{\Psi}/\tilde{\Phi}_{\|}|\ll 1$, Eq.~\eqref{eq:solution} can be simplified to
\begin{equation}
\delta \tilde{K}_{t e, n} \simeq-\frac{\omega+\omega_{*e, n}^t}{\omega+\left\langle\omega_{d e, n}\right\rangle_T}\left\langle\tilde{\Phi}_{\|}\right\rangle_T \tau F_{e}-\sum_{\nu \neq 0} \frac{\omega+\omega_{*e, n}^t}{\omega+\left\langle\omega_{d e, n}\right\rangle_T-\nu \omega_{b e}} e^{i \nu \eta_b} \mathcal{F}_b\left[\tilde{\Phi}_{\|}\right]_\nu \tau F_{e}.\label{eq:te_exact_solution}
\end{equation}
It is worth emphasizing that the leading-order term in $\delta \tilde{K}_{t e, n}$ makes no contribution to the parallel current, which corresponds to the cancellation mechanism in Eq.~\eqref{eq:te_cancellation}.
The residual trapped-electron current arising from the nonzero bounce harmonics can be expressed as
\begin{equation}
\delta \tilde{J}_{t e \|} \simeq \sum_{\nu \neq 0} q_e\left\langle v_{\|} \frac{\omega+\omega_{*e, n}^t}{\omega+\left\langle\omega_{d e, n}\right\rangle_T-\nu \omega_{b e}} e^{i \nu \eta_b} \mathcal{F}_b\left[\tilde{\Phi}_{\|}\right]_\nu \tau F_e\right\rangle_v.\label{eq:toroidal_precession_drift_contribution_to_parallel_current}
\end{equation}
As will be shown in Fig.~\ref{fig6:CTEM_GEM_currents_2}(a), this effect is an order of magnitude smaller than the passing-electron current.

\subsection{Simulation validation\label{sec3.4}}
We perform linear simulations using the GEM code~\cite{ChenY2003,ChenY2007} to diagnose the perturbed parallel currents in the electromagnetic CTEM.
GEM is a $\delta f$ particle-in-cell code that solves the gyrokinetic Vlasov-Maxwell system with gyrokinetic ions and drift-kinetic electrons.
The simulations adopt Cyclone Base Case parameters~\cite{Dimits2000,Gorler2016} and a background magnetic field with concentric circular flux surfaces.
The inverse aspect ratio is $a/R_0=0.36$, where $a$ denotes the minor radius.
The safety factor profile is $q(r)=2.52(r/a)^2-0.16(r/a)+0.86$.
The background density and temperature profiles are given by
\begin{equation}
\frac{A(r)}{A\left(r_0\right)}=\exp \left[-\kappa_A w_A \frac{a}{R_0} \tanh \left(\frac{r-r_0}{w_A a}\right)\right],
\end{equation}
where $A\in\{N_i, N_e, T_i, T_e\}$ and $r_0=0.5a$ denotes the reference radius.
The characteristic width parameter is $w_A=0.3$.
At $r=r_0$, the local parameters are $q=1.41$, $s=0.83$, $R_0/L_{n}=\kappa_{N_i}=\kappa_{N_e}=2.23$, $R_0/L_{t,e}=\kappa_{T_e}=4.46$, $R_0/L_{t,i}=\kappa_{T_i}=0$, $T_i=T_e$, $\beta_e=1\%$, and collision frequency $\nu_{\text{coll}}=0$.
The simulations employ a spatial grid of $(n_x, n_y, n_z)=(256,32,32)$ and a time step of $\Delta t=1/\Omega_{cp}$, where $\Omega_{cp}$ is the proton cyclotron frequency.
A systematic identification confirms that the observed instability is a CTEM.

\begin{figure}[htbp]
\centering
\includegraphics[width=\textwidth]{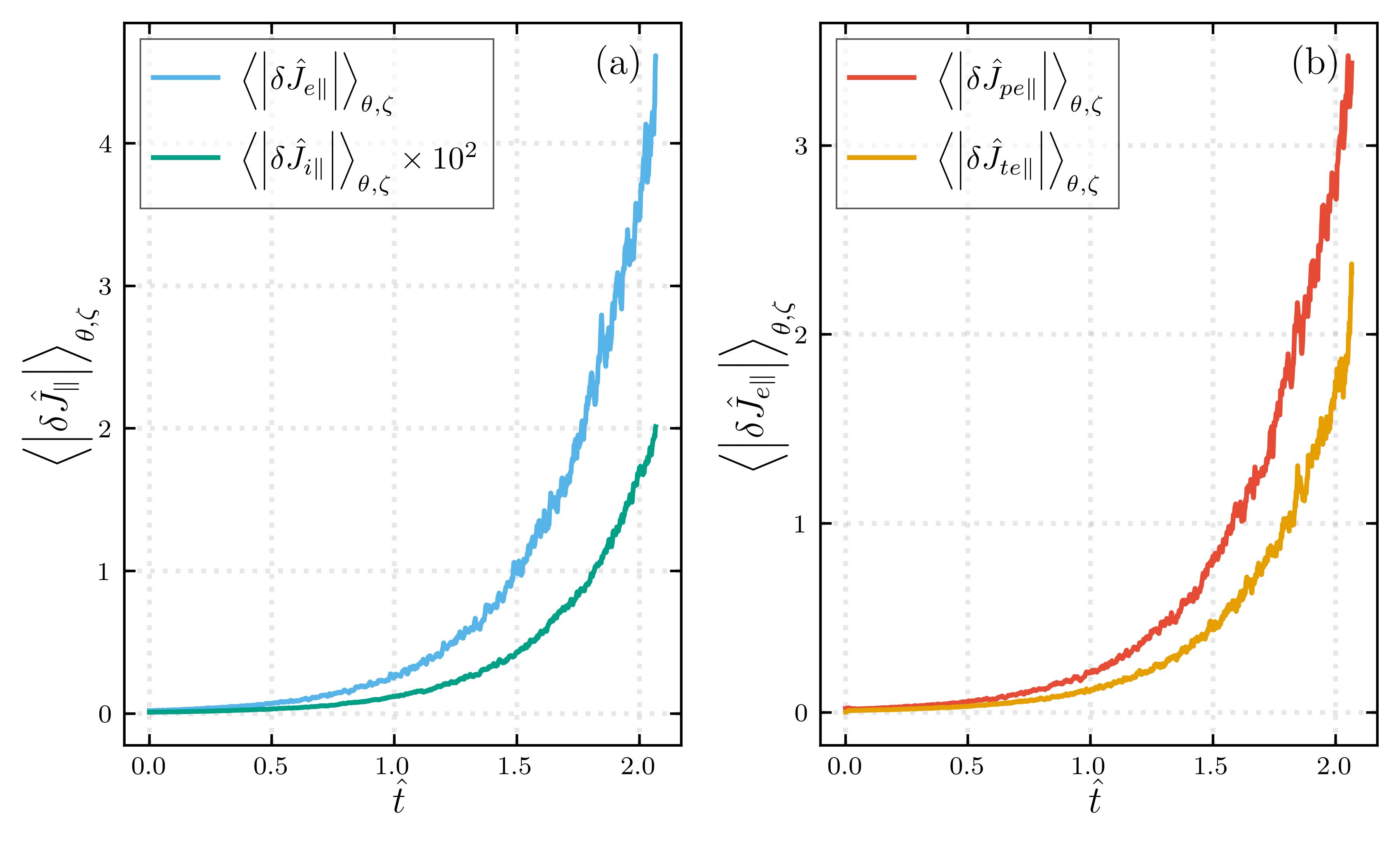}
\caption{\label{fig5:CTEM_GEM_currents_1}Time evolution of the flux-surface-averaged perturbed parallel currents in the electromagnetic CTEM. 
(a) Electron versus ion currents. 
(b) Passing- versus trapped-electron currents.}
\end{figure}

\begin{figure}[htbp]
\centering
\includegraphics[width=\textwidth]{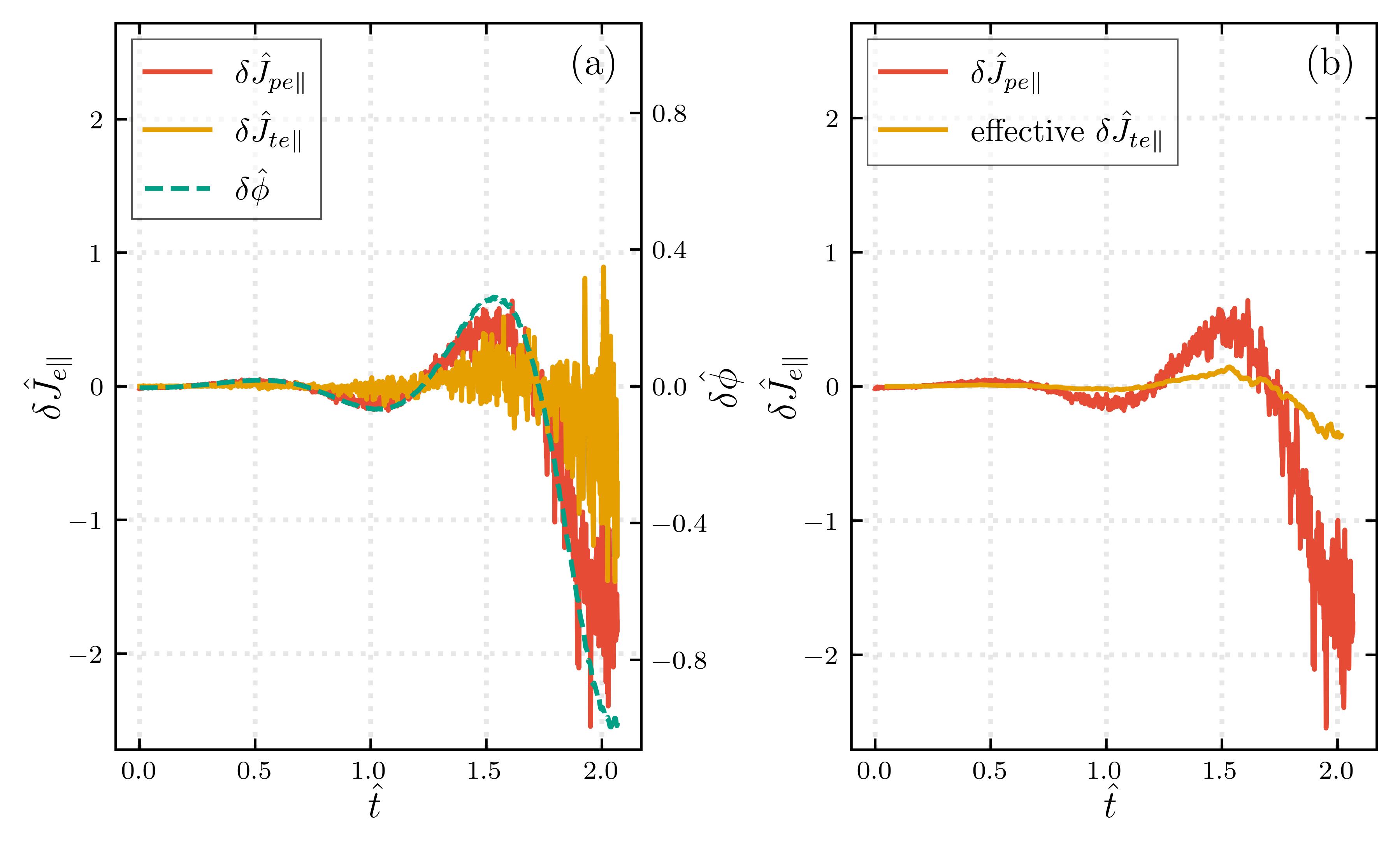}
\caption{\label{fig6:CTEM_GEM_currents_2}Time evolution of the perturbed parallel currents and electrostatic potential at the outboard midplane ($\theta = 0$) in the electromagnetic CTEM.
(a) Instantaneous passing-electron current, trapped-electron current, and electrostatic potential.
(b) Instantaneous passing-electron current versus effective (time-averaged) trapped-electron current.}
\end{figure}

Figure~\ref{fig5:CTEM_GEM_currents_1} illustrates the time evolution of the flux-surface-averaged perturbed parallel currents in the electromagnetic CTEM.
Time is normalized to the mode period, $\hat{t}=t/(2\pi/\omega_{r})$, and the parallel current is normalized as $\delta\hat{J}_{\|}=\delta J_{\|}/e N_0 v_{ti}$.
As shown in Fig.~\ref{fig5:CTEM_GEM_currents_1}(a), the ion parallel current is two orders of magnitude smaller than the electron current and is therefore negligible.
Figure~\ref{fig5:CTEM_GEM_currents_1}(b) reveals that passing and trapped electrons generate instantaneous parallel currents of comparable magnitude.
However, it is worth emphasizing that these similar instantaneous currents do not make comparable effective contributions.

To demonstrate this, we examine the local parallel currents at the outboard midplane ($\theta=0$) in Fig.~\ref{fig6:CTEM_GEM_currents_2}.
As shown in Fig.~\ref{fig6:CTEM_GEM_currents_2}(a), the trapped-electron current exhibits rapid oscillations at the bounce frequency.
Since the instantaneous currents shown in Fig.~\ref{fig6:CTEM_GEM_currents_2}(a) are already averaged over neighboring grid points, the oscillations of the trapped-electron current reflect the bounce dynamics of trapped electrons and are not numerical noise.
The effective trapped-electron current, obtained through time-averaging over the bounce period, is presented in Fig.~\ref{fig6:CTEM_GEM_currents_2}(b).
Notably, this effective current is an order of magnitude smaller than its instantaneous value, validating the cancellation mechanism in Eq.~\eqref{eq:te_cancellation}.
The parallel current in the electromagnetic CTEM is thus quantitatively confirmed to be dominated by passing electrons.
Additionally, Fig.~\ref{fig6:CTEM_GEM_currents_2}(a) reveals that the passing-electron current $\delta \hat{J}_{pe\|}$ oscillates in phase with the electrostatic potential $\delta \hat{\phi}=q_i \delta\phi/T_i$, consistent with the analytical prediction in Section~\ref{sec3.2}.

\subsection{Electromagnetic coupling analysis\label{sec3.5}}
To elucidate the physics underlying the weak electromagnetic effects of CTEMs, we analyze the coupling between the CTEM and SAW branch.
A comparison of Eqs.~\eqref{eq:vorticity_equation_in_ballooning_space}, \eqref{eq:eigenmode_vorticity}, and \eqref{eq:simplified_model} shows that $V_3 \tilde{\Psi}$ corresponds to the $\text{FLB}$ term and $\hat{\tilde{R}}_{f,pe}$ corresponds to the combined $\text{ICU}_{2,pe}+\text{MPC}_{pe}$ term, where $\text{ICU}_{2, pe}=-(1/N_0)\tau\langle[1+(\hat{\omega}_{*e,n}^t/\hat{\omega})]F_e\rangle_{v, pe}\tilde{\Phi}_{\|}$, $\text{MPC}_{pe}=(1/N_0)\tau\langle[1+(\hat{\omega}_{*e, n}^t/\hat{\omega})](\hat{\omega}_{de, n}/\hat{\omega})F_e\rangle_{v, p e} \tilde{\Psi}$.
Notably, the $\text{ICU}_{\tilde{\Psi}}$ term vanishes.
Without this term, Eq.~\eqref{eq:simplified_model} does not recover the SAW dispersion relation in the ideal MHD limit \cite{Chen2021}.
Consequently, the SAW branch is eliminated from the electromagnetic CTEM eigenmode equations at leading order.

The vanishing of the $\text{ICU}_{\tilde{\Psi}}$ term stems from the parallel particle dynamics in the CTEM regime.
For ions, the subdominant transit resonance yields $(\hat{\omega}+\hat{\omega}_{di,n})\delta\tilde{K}_{i,n}^{(0)}=(\hat{\omega}+\hat{\omega}_{*i,n}^t)J_{0i}[\tilde{\Phi}_{\|}-(\hat{\omega}_{di,n}/\hat{\omega})\tilde{\Psi}]F_i$, leading to
\begin{equation}
\begin{aligned}
&\frac{1}{N_0}\left\langle J_{0 i} \delta \tilde{K}_{i, n}^{(0)}\right\rangle_v-\frac{1}{N_0}\left\langle\left(1+\frac{\hat{\omega}_{* i, n}^t}{\hat{\omega}}\right) J_{0 i}^2 \tilde{\Phi}_{\|} F_i\right\rangle_v+\frac{1}{N_0}\left\langle J_{0 i} \frac{\hat{\omega}_{d i, n}}{\hat{\omega}} \delta \tilde{K}_{i, n}^{(0)}\right\rangle_v\\
&+\frac{1}{N_0}\left\langle\left(1+\frac{\hat{\omega}_{* i, n}^t}{\hat{\omega}}\right) J_{0 i}^2 \frac{\hat{\omega}_{d i, n}}{\hat{\omega}} \tilde{\Psi} F_i\right\rangle_v=0.
\end{aligned}
\end{equation}
This corresponds to $\text{ICU}_{1, i} + \text{ICU}_{2, i} + \text{KPC}_{i} + \text{MPC}_{i}=0$ at leading order.
For trapped electrons, as described by Eq.~\eqref{eq:te_cancellation}, the bounce dynamics yields
\begin{equation}
\begin{aligned}
&-\frac{1}{N_0}\left\langle\delta \tilde{K}_{te, n}^{(0)}\right\rangle_{v, te}-\frac{1}{N_0} \tau\left\langle\left(1+\frac{\hat{\omega}_{*e, n}^t}{\hat{\omega}}\right) \tilde{\Phi}_{\|} F_e\right\rangle_{v,te}-\frac{1}{N_0}\left\langle\frac{\hat{\omega}_{de, n}}{\hat{\omega}}\delta \tilde{K}_{te, n}^{(0)}\right\rangle_{v, te}\\
&+\frac{1}{N_0}\tau\left\langle\left(1+\frac{\hat{\omega}_{*e, n}^t}{\hat{\omega}}\right) \frac{\hat{\omega}_{d e, n}}{\hat{\omega}} \tilde{\Psi} F_e\right\rangle_{v, t e}=0.
\end{aligned}
\end{equation}
Similarly, this corresponds to $\text{ICU}_{1, te} + \text{ICU}_{2, te} + \text{KPC}_{te} + \text{MPC}_{te}=0$ at leading order.
Furthermore, the adiabatic response of passing electrons implies that $\text{ICU}_{1, pe}=0$ and $\text{KPC}_{pe}=0$.
The parallel particle dynamics reduces the gyrokinetic vorticity equation~\eqref{eq:vorticity_equation_in_ballooning_space} to
\begin{equation}
\text{FLB}+\text{ICU}_{2, pe}+\text{MPC}_{pe}=0,\label{eq:reduced_vorticity_equation_in_ctem}
\end{equation}
where the nonadiabatic charge separation term $\text{ICU}_1$ vanishes.
Thus, the effective gyrokinetic vorticity equation~\eqref{eq:reduced_vorticity_equation_in_ctem} decouples from the quasineutrality condition~\eqref{eq:quasineutrality_in_ballooning_space}, leading to the vanishing of the $\text{ICU}_{\tilde{\Psi}}$ term and the decoupling of the CTEM from the SAW branch.
In summary, in the CTEM regime, the parallel dynamics of ions and trapped electrons decouples the CTEM from the SAW branch, rendering the electromagnetic effects of CTEMs negligible.

By contrast, in the ITG regime, the dominant ion transit resonance yields $\text{ICU}_{1,i}+\text{ICU}_{2,i}+\text{KPC}_{i}+\text{MPC}_{i}\neq0$.
The quasineutrality condition is formally written as
\begin{equation}
0=\left(1+\tau\right) \tilde{\Phi}_{\|}+\left[\left(1+\frac{1}{\hat{\omega}}\right)\left(1-\Gamma_{0i}\right)+\frac{1}{\hat{\omega}}\eta_i b_i\left(\Gamma_{0i}-\Gamma_{1i}\right)\right] \tilde{\Psi}-\frac{1}{N_0}\left\langle J_{0i}\delta\tilde{K}_{i, n}\right\rangle_v,\label{eq:simplified_qn_in_itg}
\end{equation}
where the effect of trapped particles is neglected for brevity \cite{Kim1993}.
The gyrokinetic vorticity equation becomes
\begin{equation}
\text{FLB} + \text{ICU}_{1,i} + \text{ICU}_{2,i} + \text{ICU}_{2,pe} + \text{KPC}_{i} + \text{MPC}_{i} + \text{MPC}_{pe}=0.\label{eq:vn_in_itg}
\end{equation}
Given Eq.~\eqref{eq:simplified_qn_in_itg}, we note that the sum $\text{ICU}_{1,i}+\text{ICU}_{2,i}+\text{ICU}_{2,pe}$ corresponds to the combined $\text{ICU}_{\tilde{\Phi}_{\|}} + \text{ICU}_{\tilde{\Psi}}$.
A comparison of Eqs.~\eqref{eq:reduced_vorticity_equation_in_ctem} and \eqref{eq:vn_in_itg} reveals that the ion parallel dynamics is essential for the coupling between the ITG mode and the SAW branch.

\section{Conclusions\label{sec4}}
In this study, a linear gyrokinetic description of the electromagnetic CTEM is developed.
We find that the weak electromagnetic effects of CTEMs arise from particle dynamics.
Theoretical analysis reveals that in the CTEM, the perturbed parallel current is dominated by passing electrons.
For trapped electrons, the kinetic and fluid-like contributions to the current cancel at leading order.
The ion parallel current is also negligible because the ion transit resonance is subdominant.
We demonstrate that, in the gyrokinetic vorticity equation, these features of particle dynamics cause the inertia-charge uncovering term corresponding to the magnetic field perturbation to vanish, thereby eliminating the SAW branch from the electromagnetic CTEM eigenmode equations.
Consequently, the decoupling between the SAW branch and CTEM explains the negligible electromagnetic effects.

To verify our findings, we construct a reduced electromagnetic CTEM model in which the gyrokinetic vorticity equation is replaced by the parallel Amp\`ere's law.
This reduced formulation enables the selective retention of specific current contributions.
We compare two variants of this reduced model against the full gyrokinetic formulation.
The first variant retains only the passing-electron current; its agreement with the full formulation confirms that passing electrons dominate the parallel current. 
The second variant additionally includes the trapped-electron current, and its agreement confirms the cancellation mechanism of the trapped-electron current.
Furthermore, diagnostics of the parallel current in electromagnetic CTEM simulations quantitatively validate these findings.

\ack
This work was supported by the National MCF Energy R\&D Program under Grant Nos. 2022YFE03020001 and 2024YFE03230300, the National Natural Science Foundation of China under Grant No. 12375213, the Natural Science Foundation of Sichuan Province under Grant No. 2025ZNSFSC0061, the China National Nuclear Corporation 'Young Talents' Project No. 2024-QNYC-02, and the Innovation Program of Southwestern Institute of Physics (202301XWCX001).
The simulations were performed on HPC Platform of Southwestern Institute of Physics and Tianhe new generation supercomputer of National Supercomputer Center in Tianjin.

\section*{Data availability statement}
The data that support the findings of this study are available upon reasonable request from the authors.

\end{document}